\documentstyle[prl,aps,twocolumn]{revtex}

\newcommand{\oeawpreprint}[2]
{
\noindent
\begin{minipage}[t]{\textwidth}
\begin{center}
\framebox[\textwidth]{$\rule[6mm]{0mm}{0mm}$ 
\raisebox{1.3mm}{Institute for High Energy Physics of the Austrian
Academy of Sciences}}

\vspace{2mm}    \rule{\textwidth}{0.2mm}\\
\vspace{-4mm}   \rule{\textwidth}{1pt}
\mbox{ }    #1    \hfill    #2   \mbox{ }\\
\vspace{-2mm}   \rule{\textwidth}{1pt}\\
\vspace{-4.2mm} \rule{\textwidth}{0.2mm}
\end{center}
\end{minipage}

}   

\begin{document}
\input epsf
\draft
\tighten


\title{\protect\oeawpreprint{}{HEPHY-PUB 662/96, STPHY 22/96} \mbox{}\\
\mbox{}\\ \mbox{}\\ 
A sensitive test for models of Bose-Einstein correlations}

\author{H.C.\ Eggers$^{1,2}$, P.\ Lipa$^2$ and B.\ Buschbeck$^2$}
\address{$^1$Department of Physics, University of Stellenbosch,
         7600 Stellenbosch, South Africa}
\address{$^2$Institut f\"ur Hochenergiephysik,
         \"Osterreichische Akademie der Wissenschaften, 
         Nikolsdorfergasse 18, A--1050 Vienna, Austria}

\date{January 1997}

\maketitle

\begin{abstract}
{\center \bf Abstract \\}
Accurate and sensitive measurements of higher order cumulants open up
new approaches to pion interferometry. It is now possible to test
whether a given theoretical prediction can consistently match
cumulants of both second and third order. Our consistency test
utilizes a new technique combining theoretically predicted functions
with experimentally determined weights in a quasi-Monte Carlo
approach. Testing a general quantum statistics-based framework of
Bose-Einstein correlations with this technique, we find that
predictions for third order cumulants differ significantly from UA1
data. This discrepancy may point the way to more detailed dynamical
information.
\end{abstract}


\narrowtext


Pion interferometry has been a vital part of multiparticle physics for
several decades \cite{Zaj88a}. While traditionally experimental effort
in this field has centered around second order correlations, much
progress has been made recently in accurately quantifying higher order
correlations by means of so-called correlation integrals
\cite{Lip92a}, to the point where these now yield statistically
significant conclusions not only for moments but also for higher order
cumulants.  Because cumulants are so sensitive to details of the
dynamics, they represent a stringent testing ground for proposed
theoretical models.

A number of theoretical predictions for higher orders exist
\cite{Wei89a,Biy90a,Plu92a}.  In particular, Andreev, Pl\"umer and
Weiner (APW) \cite{And93a} have suggested a very general
quantum-statistical framework, based on the classical source current
formalism applied successfully in quantum optics. Its basic
assumptions are (1) a Gaussian density functional for the classical
random currents and (2) isotropy in isospin space. These two
assumptions determine all higher order correlation functions in terms
of the basic correlator, independent of the structure of the sources.
All further assumptions concern only a more detailed specification of
the space-time evolution of the sources.  Thus, the APW framework
includes as special cases more specific models of Bose-Einstein
correlations such as the GKW model \cite{Gyu79a} and the approach of
Biyajima {\it et al.\/} \cite{Biy90a}. Because the APW model is so
important, we test a simple version of its predictions below. It will
also serve as an example to show how our approach works.

While higher order cumulant measurements are valuable in their own right,
they can be used to even greater effect in {\it consistency
checks\/}: once an assumed parameterization is found to fit the second
order data, the same set of parameter values ought to fit all
predicted higher order correlations as well. Departing from tried and
tested ways, we therefore concentrate not so much on numerical values
of source parameters, but rather on utilizing their required constancy
over cumulants of different orders to test for consistency and
ultimately falsifiability of a given theoretical prediction.

Pion interferometry measures correlations in terms of {\it pair
variables\/} such as 3- or 4-momentum differences of two particles.
In the correlation integral formalism \cite{Lip92a}, this means that
the original variables in the $r$th factorial moment density
$\rho_r(\bbox{p}_1,\ldots, \bbox{p}_r)$ must be converted to
$r(r-1)/2$ (not necessarily independent) pair variables
$(q_{12},q_{13},\ldots,q_{r-1,r})$. Here, we use the specific
definition of a ``distance'' $|p_i-p_j| = \sqrt{(\bbox{p}_i -
\bbox{p}_j)^2 - (E_i - E_j)^2}$ between two phase-space points $p_i$
and $p_j$, but other pair variables and distance definitions are also
possible.  The differential moment densities as well as the
corresponding cumulant densities can be written in terms of delta
functions involving pair variables, $\delta_{12} \equiv \delta(q_{12}
- |p_1 - p_2|)$. For example, in third order we have
\[
\rho_3 (q_{12},q_{23},q_{31}) = {\textstyle\int} d\bbox{p}_1
d\bbox{p}_2 d\bbox{p}_3 \; \rho_3(\bbox{p}_1,\bbox{p}_2,\bbox{p}_3)
\delta_{12} \delta_{23} \delta_{31} \,.
\]
The pair-variable moments are measured by direct and exact sampling
prescriptions \cite{Lip92a} in terms of the measured four-momentum
differences between track pairs, $Q_{ij}$, and event averages
$\langle\cdots\rangle$,
\begin{eqnarray}
\label{acht}
\rho_2 &=& \langle {\textstyle\sum}_{i\ne j} \;\delta(q_{12} - Q_{ij})
\rangle , \\
\label{neun}
\rho_3 &=& \langle {\textstyle\sum}_{i\ne j \ne k} \; \delta(q_{12} -
Q_{ij}) \delta(q_{23} - Q_{jk}) \delta(q_{31} - Q_{ki}) \rangle\,,
\end{eqnarray}
etc. Pair-variable cumulants
\begin{eqnarray}
\label{cua}
C_2(q_{12}) &=& \rho_2(q_{12}) - \rho_1{\otimes}\rho_1(q_{12}) \,, \\
\label{cua3}
C_3(q_{12},q_{23},q_{31}) &=& \rho_3 - {\textstyle\sum}_{(3)}
\rho_2{\otimes}\rho_1 + 2 \rho_1{\otimes}\rho_1{\otimes}\rho_1 \,,
\end{eqnarray}
are similarly measured in exact prescriptions.  Mixed terms such as
$\rho_1{\otimes}\rho_1$ require multiple event averages in terms of
distances $Q_{ij}^{ab} \equiv | p_i^a - p_j^b |$ between particles
$p_i^a$ and $p_j^b$ taken from different events $a$ and $b$,
\begin{equation}
\label{cuo}
\rho_1{\otimes}\rho_1(q_{12}) = \langle \langle {\textstyle\sum}_{i,
j} \delta(q_{12} - Q_{ij}^{ab}) \rangle \rangle
\end{equation}
and so on (see Refs.\cite{Lip92a} for details).

These experimentally measured cumulants are, after normalization, to
be compared to the corresponding theoretical predictions.  The APW
normalized cumulant predictions are built up from normalized
correlators $d_{ij}$, the on-shell Fourier transforms of the
space-time classical current correlators.  The specific
parameterizations of the correlator we shall be testing are, in terms
of the 4-momentum difference $q_{ij}$,
\begin{eqnarray}
\label{cuh}
{\rm Gaussian{:}\ \ \ \ \ \ } d_{ij} &=& \exp(-r^2 q_{ij}^2) \,,
\mbox{\ \ \ \ \ \ } \\
\label{cui}
{\rm exponential{:}\ \ \ \ \ \ } d_{ij} &=& \exp(-r q_{ij}) \,,
\mbox{\ \ \ \ \ \ } \\
\label{cuj}
{\rm power\ law{:}\ \ \ \ \ \ } d_{ij} &=& q_{ij}^{-\alpha} \,.
\mbox{\ \ \ \ \ \ }
\end{eqnarray}
For constant chaoticity $\lambda$ and real-valued currents, APW
predict the second and third-order normalized cumulants $k_2^{\rm
th}(q_{12})$ and $k_3^{\rm th}(q_{12},q_{23},q_{31})$ to have the form
\cite{And93a}
\begin{eqnarray}
\label{cue}
k_2^{\rm th} \equiv {C_2 \over \rho_1{\otimes}\rho_1} &=& 2\lambda
   (1-\lambda )d_{12} + \lambda^2 d_{12}^2 \,, \\
\label{cuf}
k_3^{\rm th} \equiv {C_3 \over \rho_1{\otimes}\rho_1{\otimes}\rho_1}
   &=& 2\lambda^2(1-\lambda) [ d_{12}d_{23} + d_{23}d_{31} +
   d_{31}d_{12} ]\nonumber \\ && + 2\lambda^3 d_{12} d_{23} d_{31} \,,
\end{eqnarray}
with any of the above $d_{ij}$ parameterizations.

A comparison of sample cumulants (\ref{cua})--(\ref{cua3}) with the
theoretical APW cumulants in (\ref{cue})--(\ref{cuf}) encounters two
basic difficulties: projection and normalization. In dealing with
these, we are led to the ``Monte Karli'' prescription outlined below.

Statistical sampling limitations render a fully multidimensional
measurement of cumulants such as $C_3(q_{12},q_{23},q_{31})$
impossible, necessitating projections onto a single variable. Such
projections must of course then be applied to theoretical predictions
such as (\ref{cuf}) also. However, the complicated combinatorics of
$d_{ij}$'s entering the cumulants make it generally impossible to give
analytical formulae in terms of any one combined variable (such as
$S=q_{12}+q_{23}+q_{31}$) without making further approximations.
While two such approximations, leading to simple third order formulae,
have been used previously by us and other groups
\cite{E1,DELPHI-95a,NA22_95a}, conclusions based on them are weak
because the effects of making such approximations are seldom
quantifiable.

The difficulty in normalization arises because experimental and
theoretical procedures differ. Theo\-re\-ti\-cal cumulants are
normalized fully differentially with $\rho_1{\otimes}\rho_1(q_{12})$
and $\rho_1{\otimes}\rho_1{\otimes}\rho_1(q_{12},q_{23},q_{31})$
respectively.  Such normalizations assume essentially perfect
measurement accuracy for the $q_{ij}$'s and infinite sample size.
Experimentally, however, one always measures {\it unnormalized\/}
cumulants over some phase-space region $\Omega$ and then divides by an
uncorrelated reference sample integrated over the same $\Omega$. For
example, in second order one actually measures
\begin{equation}
\label{cum}
\Delta K_2^{\rm ex}(\Omega) = {\int_\Omega C_2^{\rm ex}(q)\, dq \over
\int_\Omega \rho_1{\otimes}\rho_1^{\rm ex}(q)\, dq} \,,
\end{equation}
which approaches the differential $k_2^{\rm th}(q)$ or the
bin-integrated normalized theoretical cumulant
\begin{equation}
\label{cuu}
\Delta K_2^{\rm th}(\Omega) 
  = {1\over\Omega} \int_\Omega k_2^{\rm th}(q) \; dq 
\end{equation}
only in the limit $\Omega\to 0$ and may significantly differ from
$\Delta K_2^{\rm th}$ for larger $\Omega$.  Non-stationary
single-particle distributions can significantly distort results,
especially when they vary strongly within a given bin, as they do in
the case of momentum differences.  The problem is exacerbated in
higher orders.

``Monte Karli'' (MK) integration provides a solution to both these
problems  \cite{E1}. It is conceptually quite simple: Since $C_2(q) =
k_2(q) \, \rho_1{\otimes}\rho_1(q)$ we can, in analogy to $\Delta
K_2^{\rm ex}$, define an MK normalized cumulant
\begin{equation}
\label{cun}
\Delta K_2^{\rm MK}(\Omega) \equiv {\int_\Omega k_2^{\rm th}(q) \;
\rho_1{\otimes}\rho_1^{\rm ex}(q) \, dq \over \int_\Omega
\rho_1{\otimes}\rho_1^{\rm ex}(q)\, dq} \equiv {\int_\Omega C_2^{\rm
MK}(q) \; dq \over \int_\Omega \rho_1{\otimes}\rho_1^{\rm ex}(q)\, dq}
\,,
\end{equation}
where $k_2^{\rm th}$ is a function supplied by theory while
$\rho_1{\otimes}\rho_1^{\rm ex}$ is taken directly from experiment.
Thus $\Delta K_2^{\rm MK}$ by construction satisfies both the
theoretical function $k_2^{\rm th}$ and the experimental one-particle
distributions. Drawing $k_2^{\rm th}$ into the event averages of the
correlation integral prescription (\ref{cuo}), the unnormalized MK
cumulant becomes
\begin{equation}
\label{cumk2}
C_2^{\rm MK}(q) = \langle \langle {\textstyle\sum_{i, j}} \delta(q -
Q_{ij}^{ab}) k_2^{\rm th}(Q_{ij}^{ab}) \rangle \rangle ,
\end{equation}
so that the argument of the theoretical $k_2^{\rm th}$ is now an
experimental particle pair distance, obtained by event mixing.
Prescriptions such as (\ref{cumk2}) and (\ref{cusy}) below amount to
Monte Carlo-style sampling of the {\it theoretical function\/} weighted
by the actual {\it experimental particle distributions\/}, thereby
earning Monte Karli its name.  Integrated over the bin domain $\Omega$,
the normalized MK cumulant reads
\begin{equation}
\label{cuss}
\Delta K_2^{\rm MK}(\Omega) = { \langle \langle \sum_{i, j}
\int_\Omega dq \, \delta(q - Q_{ij}^{ab}) k_2^{\rm th}(Q_{ij}^{ab})
\rangle \rangle \over \langle \langle \sum_{i, j} \int_\Omega dq \,
\delta(q - Q_{ij}^{ab}) \rangle \rangle} ,
\end{equation}
where binning of track pairs is represented by $\int_\Omega dq\,
\delta(q - Q_{ij}^{ab})$: whenever a pair's $Q_{ij}^{ab}$ falls within
$\Omega$, the numerator counts $k_2^{\rm th}(Q_{ij}^{ab})$ while the
denominator counts 1.

Similarly, the unnormalized third-order Monte Karli cumulant $C_3^{\rm
MK}(q)$ can be written in terms of $k_3^{\rm th}$ times
$\rho_1{\otimes}\rho_1{\otimes}\rho_1^{\rm ex}$.  For the projection
onto the single variable $q$, different ``topologies'' exist which
quantify the size of a given cluster of $r$ particles in phase space
\cite{Lip92a}. The ``GHP max'' prescription used here bins cluster
sizes according to the maximum 4-momentum difference.  Inserting this
into the correlation integral prescription, one finds
\begin{eqnarray}
\label{cusy}
C_3^{\rm MK}(q) &=& \Bigl\langle \!\! \Bigl\langle \!\!  \Bigl\langle
\sum_{i, j, k} \delta\bigl[q - \max(Q_{ij},Q_{jk},Q_{ki}) \bigr]
\nonumber \\ &&
k_3^{\rm th}(Q_{ij}^{ab},Q_{jk}^{bc},Q_{ki}^{ca}) \Bigr\rangle \!\!
\Bigr\rangle \!\! \Bigr\rangle  ,
\end{eqnarray}
and the normalized binned MK cumulant becomes
\begin{eqnarray}
\label{cust}
&&{} \Delta K_3^{\rm MK}(\Omega) = \\ && { \langle \langle \langle
\sum_{i, j, k} \int_\Omega dq\, \delta[q - \max(\cdots)] \, k_3^{\rm
th}(Q_{ij}^{ab},Q_{jk}^{bc},Q_{ki}^{ca}) \, \rangle \rangle \rangle
\over \langle \langle \langle \sum_{i, j, k} \int_\Omega dq\, \delta[q
- \max(Q_{ij}^{ab},Q_{jk}^{bc},Q_{ki}^{ca})] \, \rangle \rangle
\rangle \nonumber } .
\end{eqnarray}
Multiple event averages are handled efficiently within a ``reduced
event mixing'' algorithm using unbiased estimators, cf.\ Eggers and
Lipa in \cite{Lip92a}.  Because the identical projection is applied to
both experimental and theoretical cumulants, their comparison is not
subject to any approximation or limitation.  Clearly, the MK procedure
can be generalized to all possible moments and cumulants,
independently of variable or integration topology.

We have measured second- and third-order normalized cumulants using a
sample of around 160,000 non-single-diffractive events at 630 GeV/c
recorded in the UA1 central detector. Only vertex-associated charged
tracks with transverse momentum $p_\perp \geq 0.15$ GeV/c,
pseudorapidity $|\eta | \leq 3$ and good measurement quality were
used. We restricted our analysis to the azimuthal angle region
$45^\circ \leq |\phi| \leq 135^\circ$ because there acceptance
corrections are small enough to be safely neglected. To calculate the
four momentum difference $q$, all particles were assumed to be pions.
The resolution $\Delta q \simeq 8$~MeV, estimated from the errors of
track fits, is approximately constant over the whole range $q \leq
1$~GeV. For further details of the detector and pair selection
criteria, see Ref.\cite{UA1-93a}.

In the following, we check the APW formulae (\ref{cuh})--(\ref{cuf})
for consistency with the data. The procedure is to find numerical
values for $r$ (or $\alpha$) and $\lambda$ by fitting the model's
second-order prediction (\ref{cue}) for $k_2$ to the experimentally
measured differential cumulant. These best-fit values are ported into
the third-order prediction (\ref{cuf}) which are then compared to
third-order data using (\ref{cust}).

In Figure 1, we show the second-order cumulant of like-sign particles
$\Delta K_2 = (\int \rho_2 / \int \rho_1{\otimes}\rho_1 )$ $-1$, where
numerator and denominator are integrals over bins spaced
logarithmically between $q = 1$ GeV and 30 MeV\footnote{
It should be remarked that previous work has shown the utility of
using logarithmic rather than linear binning: much of what is
interesting in correlations happens at small $q$, and this region is
probed better by using logarithmic bins.}.
Fits to the data were performed using the three parameterizations
(\ref{cuh})--(\ref{cuj}) in the APW form (\ref{cue}). All fits shown
include, besides the free parameters $\lambda$ and $r$ (or $\alpha$),
an overall additive constant $A$ as free parameter.
Such additive constants are necessary for our choice of normalization
because of the non-Poissonian nature of UA1 data, which leads to
nonvanishing cumulants at large $q$.
Best-fit parameter values obtained were $\lambda = 0.050 \pm 0.005$,
$\alpha = 0.64 \pm 0.03$ ($A = 0.269 \pm 0.012$) for the APW power
law, and $\lambda = 0.59 \pm 0.04$, $r = 1.16 \pm 0.02$ fm ($A = 0.385
\pm 0.003$) for the APW exponential.  The Gaussian parameters are
$\lambda = 0.23\pm 0.01$, $r = 0.81 \pm 0.02$ fm ($A = 0.396\pm
0.002$).  Goodness-of-fits were $\chi^2/{\rm NDF} = 3.34,\ 6.65$ and
20.0 for APW power, exponential and Gaussian respectively.
\begin{figure}
\centerline{ \epsfysize=100mm \epsfbox{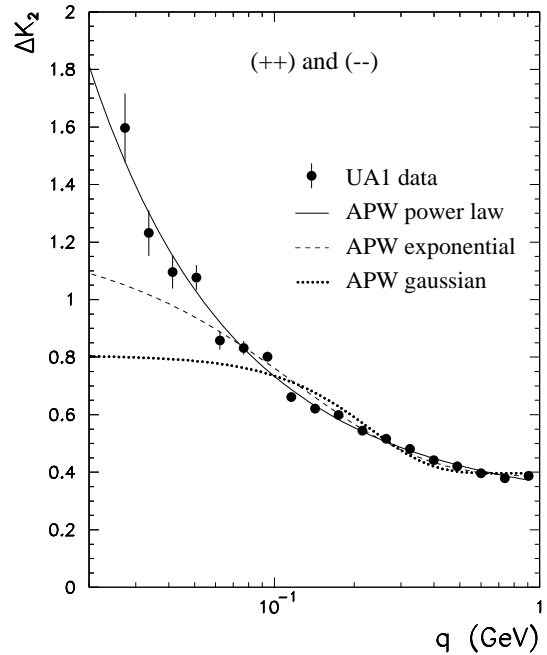} }

\caption{Second order like-sign UA1 cumulant with fits using various
parameterizations for $d_{12}$ in formula (\ref{cue}). }
\label{fig1}
\end{figure}

For $\Delta K_3$, we have performed three separate consistency checks,
two based on approximations (shown in \cite{E1}), the third involving
the Monte Karli method (shown below) which is exact. In
Ref.\ \cite{E1}, the APW predictions for $\Delta K_3$ based on the two
approximations were found to differ substantially from the data.

In Figure 2, the results of implementing the MK prescription are
shown, using the GHP max topology and the parameterizations
(\ref{cui}) and (\ref{cuj}). Parameter values used for the respective
power law and exponential MK points were taken from the fit to $\Delta
K_2$ of Figure 1. (The effect of finite binning was checked by
inserting these parameter values back into the MK formulae for $\Delta
K_2$ and finding good agreement between UA1 data and MK output.)
Again, all MK points shown are determined only up to an additive
constant, so that they may be shifted up and down in unison.
It is clear, though, that the shape of the measured third-order
cumulant differs significantly from that predicted by the APW formulae
and $\Delta K_2$ parameter values.  Similar discrepancies were found
when using other topologies such as the GHP sum variable \cite{E1}.
\begin{figure}
\centerline{ \epsfysize=100mm \epsfbox{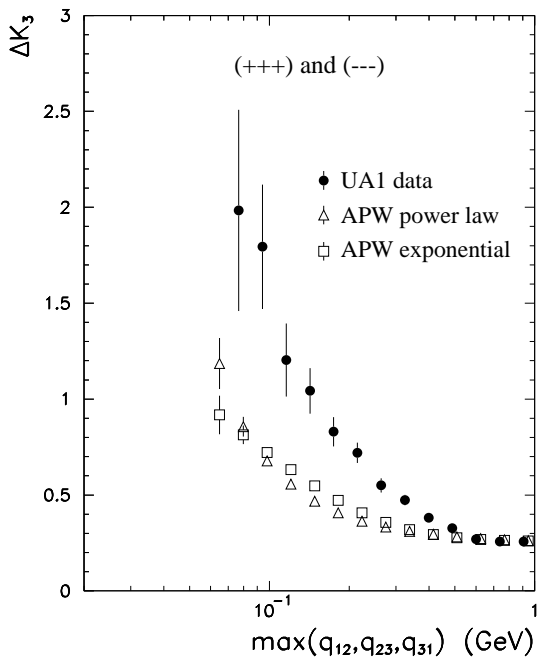} }

\caption{ UA1 third-order ``GHP max'' cumulant data, together with APW
predictions based on Monte Karli integration and parameter values
taken from $\Delta K_2$. }
\label{fig3}
\end{figure}

This result does not contradict an earlier UA1 paper \cite{UA1-93a}
whose higher order moment data was claimed in \cite{Plu92a} to be in
agreement with the APW formalism. The reason is that moments are
numerically dominated by a combinatoric background of lower orders, so
that comparisons based on moments are not very sensitive. Higher-order
cumulants are a much more sensitive test of the theory, as we have
just shown.

We have considered the following sources of errors:
\begin{itemize}
\item
No corrections for Coulomb repulsion were included. However, $\Delta
K_3$ data rise more strongly than theoretical predictions even for
large $q$ (several hundred MeV) where Coulomb repulsion is not
expected to be important.

\item
We have checked through additional small-$q$ cuts that possible
residual track mismatching (multiple counting of tracks
\cite{UA1-93a}, \cite{Lipphd}) does not explain the discrepancy
between predicted and observed $\Delta K_3$.

\item
Our sample consists of 15\% unidentified kaons and protons which were
treated as pions in our analysis. This impurity is expected to weaken
the correlations. We have repeated the analysis with a low-momentum
($|\bbox{p}|\leq 0.6$~GeV/c) sample of particles by using the
information on $dE/dx$ to get an almost pure sample of pions. No
reduction of the discrepancy between the predicted and observed
$\Delta K_3$ was observed.

\item
We have checked whether the restriction to the azimuthal angle region
$45^\circ \leq |\phi| \leq 135^\circ$ distorts $\Delta K_2$ or $\Delta
K_3$. In the region $q < 1$ GeV, no distortion is observed.
\end{itemize}

Nevertheless, we do not regard our result as a failure of the APW
formalism per se. One caveat relates to the structure of the UA1
multiplicity distribution which is non-Poissonian (i.e.\ the cumulants
are nonzero at large $q$). The use of an additive constant in the
cumulants is only an approximate remedy which can be improved
substantially \cite{Lip96a}.

Furthermore, we assumed the chaoticity $\lambda$ to be {\it
momentum-independent\/} and the source currents {\it real-valued}, as
has been done implicitly in all previous experimental work
known to us.  More general versions of the APW theory have yet to be
tested. Indeed, we hope that with our sensitive new testing methods the
role of such hitherto inaccessible dynamical information can be
investigated.  Beyond such extensions there may lurk a non-Gaussian
source current.


\acknowledgements We thank I.\ Andreev, A.\ Bartl, J.\ D.\ Bjorken,
T. Cs\"org\H{o} and R.\ Weiner for useful discussions.  This work was
supported in part by the Austrian Fonds zur F\"orderung der
wissenschaftlichen Forschung (FWF) by means of a Lise-Meitner
fellowship (HCE), by the Foundation for Research Development and by an
APART fellowship of the Austrian Academy of Sciences (PL).

\end{document}